\documentclass[12pt]{article}

\usepackage{amsmath,amssymb,epic,epsfig,amscd}
\usepackage{color}
\usepackage{graphicx}

\title{From  Yang-Baxter maps \\ to integrable quad maps and recurrences}
\author{B. Grammaticos \\ IMNC, Universit\'es Paris VII \& XI, CNRS  UMR 8165 \\  B\^at. 440, 91406 Orsay, France \and A. Ramani \\Centre de Physique Th\'eorique, Ecole Polytechnique, CNRS UMR 7644  \\ 91128 Palaiseau, France   \and C-M. Viallet \\ LPTHE, Universit\'e Pierre et Marie Curie, CNRS UMR 7589  \\ 4 Place Jussieu, 75252 Paris Cedex 05, France}

\begin{document}
\maketitle

\begin{abstract}
Starting from known solutions of the functional Yang-Baxter equations,
we exhibit Miura type of transformations leading to various known
integrable quad equations.  We then construct, from the same list of
Yang-Baxter maps, a series of non-autonomous solvable recurrences of
order two.
\end{abstract}

\section{Introduction}

A number of discrete integrable systems have been described in the
literature in the last decades. They are mostly discrete versions of
known integrable differential equations and partial differential
equations, retaining fundamental features of integrability, as for
example the existence of a sufficient number of invariant tori and
hierarchies of associated integrable equations. Important
classification results~\cite{Sa01} have been obtained, both for the
ordinary difference equations case (in particular integrable
maps~\cite{QuRoTh89} and discrete Painlev\'e
equations~\cite{RaGrHi91}) and for the partial difference equations
(integrable lattice equations and Yang-Baxter maps among
others~\cite{AdBoSu03,AdBoSu07,PaSuToVe10}).

We show that there exist bridges between different systems.  We
examine two such situations. Starting from Yang-Baxter maps, given as
lattice equations on a quad graph with variables on the bonds, we
first construct (section~\ref{miura}), via Miura type of
transformations, quad maps with variables on the vertices,
establishing a correspondence between the classifications
of~\cite{AdBoSu03} and \cite{AdBoSu07}.  We then construct, from the
same list of Yang-Baxter maps a series of solvable recurrences
(section~\ref{yb2rec}).

\section{The setting}

The underlying geometry behind the models is a two dimensional square
lattice with variables associated to the edges and/or to the
vertices. Each elementary cell is labelled by a pair of
integers $(m,n)$, which is the pair of coordinates of the lower left
corner of the cell. Whenever possible we will use the standard
notation ``bar'' and ``tilde'' to indicate shifts in the $n$ and $m$
direction respectively i.e.  $w\equiv w(m,n)$, $\bar w\equiv
w(m,n+1)$, $\tilde w\equiv w(m+1,n)$ and $\skew{3.5}\tilde{\bar
  w}\equiv w(m+1,n+1)$. The elementary cell is depicted in Figure 1.

\begin{center}
\setlength{\unitlength}{0.0007in}
\begin{picture}(3482,2000)(0,-10)
\put(1350,1883){\circle*{91}}
\put(2250,983){\circle*{90}}
\put(450,983){\circle*{90}}
\put(1350,83){\circle*{90}}
\put(100,900){\makebox(0,0)[lb]{$x$}}
\put(2380,900){\makebox(0,0)[lb]{$u$}}
\put(1250,-200){\makebox(0,0)[lb]{$y$}}
\put(1250,2083){\makebox(0,0)[lb]{$v$}}
\drawline(450,1883)(450,83)
\drawline(450,1883)(2250,1883)
\drawline(2250,1883)(2250,83)
\drawline(450,83)(2250,83)
\end{picture}
\\
\bigskip
\hskip -1truecm Figure 1: \em{Original  cell}
\end{center}

The variables ($x,y,u,v$) are assigned to the bonds as shown
in Figure~1. The models are defined by giving a $2 \rightarrow 2$ map
$(x,y) \rightarrow (u,v)$, verifying the functional Yang-Baxter
equations (aka Yang-Baxter maps).

We start from the models described in~\cite{AdBoSu04,PaSuToVe10} and
use the same labelling, namely the ``F'' and ``H'' families, which we
reproduce here:
 
\begin{align}
\label{F1}\tag{$F_{\rm{I}}$}
  u= \alpha y\; P,\quad &
  v= \beta x \; P,\quad &
  P=& \frac{(1-\beta)x+\beta-\alpha+(\alpha-1)y}
           {\beta(1-\alpha)x+(\alpha-\beta)yx+\alpha(\beta-1)y}, \\
\label{F2}\tag{$F_{\rm II}$}
  u= \frac{y}{\alpha} P, \quad &
  v= \frac{x}{\beta} P,\quad &
  P=& \frac{\alpha x-\beta y+\beta-\alpha}{x-y}, \\
\label{F3}\tag{$F_{\rm III}$}
  u= \frac{y}{\alpha} P,\quad &
  v= \frac{x}{\beta} P,\quad & 
  P=& \frac{\alpha x-\beta y}{x-y}, \\
\label{F4}\tag{$F_{\rm IV}$}
  u= y \; P,\quad &
  v= x\; P,\quad &
  P =& 1+\frac{\beta-\alpha}{x-y}, \\
\label{F5}\tag{$F_{\rm V}$}
  u= y+P,\quad &
  v= x+P,\quad &
  P=& \frac{\alpha-\beta}{x-y},
\end{align}

and 
\begin{align}
\label{H1}\tag{$H_{\rm I}$}
  u= y \; Q^{-1},\quad & v= x \; Q,\quad & Q=&
  \frac{(1-\beta)xy+(\beta-\alpha)y+\beta(\alpha-1)}
       {(1-\alpha)xy+(\alpha-\beta)x+\alpha(\beta-1)},\\
\label{H2}\tag{$H_{\rm II}$}
  u= y \; Q^{-1},\quad & v= x \; Q,\quad & Q= &\frac{\alpha +
    (\beta-\alpha) y - \beta xy}{\beta + (\alpha-\beta)x -\alpha
    xy},\\
\label{H3A}\tag{$H_{\rm III}^A$}
  u= \frac{y}{\alpha} Q,\quad &
  v= \frac{x}{\beta} Q,\quad &
 Q= &\frac{\alpha x + \beta y}{x+y},\\
 \label{H3B}\tag{$H_{\rm III}^B$}
  u= y \; Q^{-1},\quad &
  v=  x \; Q,\quad &
 Q=& \frac{\alpha xy + 1}{\beta xy+1},\\
\label{H5}\tag{$H_{\rm V}$}
  u= y-P,\quad &
  v= x+P,\quad &
  P=& \frac{\alpha-\beta}{x+y}.
\end{align}

We will discard the cases $H_{\rm III^A}$ and $H_{\rm V}$ which are
equivalent respectively to $F_{\rm III}$ and $F_{\rm V}$, through the
change $u \rightarrow -u$ and $v \rightarrow -v$.

In section~\ref{miura} we establish a correspondence between the
original models and known integrable ``quad equation'', that is to say
partial difference equations, which happen to fall in a classification
given by the same authors~\cite{AdBoSu07}.

In section~\ref{yb2rec} we construct from the same family of models a
number of solvable ordinary difference equations.

\section{A  correspondence via Miura  transforms}
\label{miura}

 We will bring in vertex variables
 ($w,\tilde{w},\bar{w},\tilde{\bar{w}})$ (see Figure~2), subject to
 some local condition, via a Miura type of transformation of the
 original variables ($x,y,u,v$).

\begin{center}
\setlength{\unitlength}{0.0007in}
\begin{picture}(3482,2000)(0,-10)
\put(450,1883){\circle{91}}
\put(1350,1883){\circle*{91}}
\put(2250,1883){\circle{91}}
\put(2250,983){\circle*{90}}
\put(450,983){\circle*{90}}
\put(2250,83){\circle{90}}
\put(1350,83){\circle*{90}}
\put(450,83){\circle{90}}
\put(120,1950){\makebox(0,0)[lb]{$\tilde{w}$}}
\put(2380,0){\makebox(0,0)[lb]{$\bar{w}$}}
\put(2380,1950){\makebox(0,0)[lb]{$\tilde{\bar{w}}$}}
\put(120,0){\makebox(0,0)[lb]{$w$}}
\put(100,900){\makebox(0,0)[lb]{$x$}}
\put(2380,900){\makebox(0,0)[lb]{$u$}}
\put(1250,-200){\makebox(0,0)[lb]{$y$}}
\put(1250,2083){\makebox(0,0)[lb]{$v$}}
\drawline(450,1883)(450,83)
\drawline(450,1883)(2250,1883)
\drawline(2250,1883)(2250,83)
\drawline(450,83)(2250,83)
\end{picture}
\\
\bigskip
\hskip -1.5 truecm Figure 2: {\em{Decorated  cell}} 
\end{center}

\subsection{ $F_{\rm I}$}
The first case we  examine is $F_{\rm I}$:
\begin{equation}
u=\alpha\; y\; P,\qquad v=\beta\; x\; P,\qquad
P={(1-\beta)x+\beta-\alpha+(\alpha-1)y\over
  \beta(1-\alpha)x+(\alpha-\beta)xy+\alpha(\beta-1)y}\label{fi1}
\end{equation}
Define vertex variables ($w,\tilde{w},\bar{w},\tilde{\bar{w}})$ via
the Miura transformation:
\begin{equation}
x=p\; w\tilde w,\quad u=p\; \bar w\skew{3.5}\tilde{\bar w},\quad v=q\; w\bar
w,\quad y=q\; \tilde w\skew{3.5}\tilde{\bar w}\label{fi2}
\end{equation}
It is  straightforward to eliminate the $x, y, v, u$ variables using
Eq.~(\ref{fi1}) and obtain an equation for $w$ alone.  Setting $\alpha=p^2$
and $\beta=q^2$, we obtain exactly equation A2 of \cite{AdBoSu07}:
 \begin{equation}
(q^2-p^2)(w\tilde w\bar w\skew{3.5}\tilde{\bar
  w}+1)+p(1-q^2)(w\tilde w+\bar w\skew{3.5}\tilde{\bar
  w})-q(1-p^2)(w\bar w+\tilde w\skew{3.5}\tilde{\bar
  w})=0\label{fi3}
\end{equation} in canonical form.

\subsection{$F_{\rm II}$}
In the case of $F_{\rm II}$
\begin{equation}
u={y\over\alpha}P,\qquad v={x\over\beta}P,\qquad P={\alpha x-\beta
  y+\beta-\alpha\over x-y}\label{fii1}
\end{equation} 
we introduce the Miura transformation:
\begin{equation}
x=p\; w\tilde w,\quad u=p\; \bar w\skew{3.5}\tilde{\bar w},\quad v=q\; w\bar
w,\quad y=q\; \tilde w\skew{3.5}\tilde{\bar w}\label{fii2}
\end{equation} with 
$q^2/p^2=\alpha/\beta$.  We get
\begin{equation}
q\; (w\tilde w+\bar w\skew{3.5}\tilde{\bar w})-p\; (w\bar w+\tilde
w\skew{3.5}\tilde{\bar w})+{p\over q}-{q\over p}=0\label{fii3}
\end{equation}
which is equation H3 of \cite{AdBoSu07}.

\subsection{$F_{\rm III}$}
For system $F_{\rm III}$ we start from
\begin{equation}
u={y\over\alpha}P,\qquad v={x\over\beta}P,\qquad P={\alpha x-\beta
  y\over x-y}\label{fiii1}
\end{equation} and introduce the Miura transformation
\begin{equation}
x=p\; w\tilde w,\quad u=p\; \bar w\skew{3.5}\tilde{\bar w},\quad v=q\; w\bar
w,\quad y=q\; \tilde w\skew{3.5}\tilde{\bar w}\label{fiii2}
\end{equation}
Substituting into the two equations we find the condition
$q^2/p^2=\alpha/\beta$ whereupon $w$ satisfies the equation
\begin{equation}
q\; (w\tilde w+\bar w\skew{3.5}\tilde{\bar w})-p\; (w\bar w+\tilde
w\skew{3.5}\tilde{\bar w})=0\label{fiii3}
\end{equation} 
This is precisely the H3 equation of \cite{AdBoSu07} for the special value
$\delta=0$. A more familiar interpretation of this equation is
possible. Indeed solving for $\skew{3.5}\tilde{\bar w}$ we obtain
\begin{equation}
\skew{3.5}\tilde{\bar w}=w\; {q\tilde w-p\bar w\over p\tilde w-q\bar
  w}\label{fiii4}
\end{equation} which is just the potential form of the discrete
modified KdV equation.

\subsection{ $F_{\rm IV}$}
The defining equation is
\begin{equation}
u=y\; P,\qquad v=x\; P,\qquad P=1-{\alpha-\beta\over x-y}\label{fiv1}
\end{equation}
Introduce the Miura transformations
\begin{equation}
x=p+w+\tilde w,\quad u=p+\bar w+\skew{3.5}\tilde{\bar w},\quad
v=q+w+\bar w,\quad y=q+\tilde w+\skew{3.5}\tilde{\bar w}\label{fiv2}
\end{equation}
where $\beta-\alpha=2(p-q)$.  Eliminating the $x,y,v,u$ variables
using Eq.~(\ref{fiv1}) we obtain an equation for $w$ alone
\begin{equation}
(\skew{3.5}\tilde{\bar w}-w)(\tilde w-\bar
w)+(p-q)(\skew{3.5}\tilde{\bar w}+\tilde w+\bar
w+w)+p^2-q^2=0\label{fiv3}
\end{equation} Equation (\ref{fiv3}) is precisely
equation H2 of~\cite{AdBoSu07}.

\subsection{ $F_{\rm V}$}
The defining  equation is
\begin{equation}
u=y+P,\qquad v=x+P,\qquad P={\alpha-\beta\over x-y}\label{fv1}
\end{equation}
We introduce the $w$ variable by
\begin{equation}
x=w+\tilde w,\quad u=\bar w+\skew{3.5}\tilde{\bar w},\quad v=w+\bar
w,\quad y=\tilde w+\skew{3.5}\tilde{\bar w}. \label{fv2}
\end{equation}
 It is straightforward to eliminate the $x,y,v,u$ variables using
 Eq.~(\ref{fv1}) and obtain an equation for $w$ alone. We find
\begin{equation}
(\skew{3.5}\tilde{\bar w}-w)(\bar w-\tilde
w)=\alpha-\beta\label{fv3}
\end{equation} 
which is the potential form of the discrete KdV equation that is to
say equation H1 of~\cite{AdBoSu07}.

Next we turn to the study of the $H$ family.

\subsection{ $H_{\rm I}$}
The first case we examine is $H_{\rm I}$
\begin{equation}
u=y\; Q^{-1},\qquad v=x\; Q,\qquad
Q={(1-\beta)xy+(\beta-\alpha)y+\beta(\alpha-1)\over(1-\alpha)xy+(\alpha-\beta)x+\alpha(\beta-1)}\label{hi1}
\end{equation}
The Miura transformation is 
\begin{equation}
x=p\; w\tilde w,\quad u=p\; (\bar w\skew{3.5}\tilde{\bar w})^{-1},\quad
v=q\; w\bar w,\quad y=q\; (\tilde w\skew{3.5}\tilde{\bar
  w})^{-1}\label{hi2}
\end{equation} A direct substitution into Eq.~(\ref{hi1}) leads
to an equation which does not have the canonical A2 form. However
introducing  $\alpha=p^2$ and $\beta=q^2$ we obtain
 \begin{equation}
(q^2-p^2)\; (w\tilde w\bar w\skew{3.5}\tilde{\bar w}+1)+p(1-q^2)\;
   (w\tilde w+\bar w\skew{3.5}\tilde{\bar w})-q\; (1-p^2)(w\bar
   w+\tilde w\skew{3.5}\tilde{\bar w})=0\label{hi3}
\end{equation} which is A2 in canonical form.

\subsection{ $H_{\rm II}$}
The defining relation is
\begin{equation}
u=y\; Q^{-1},\qquad v=x\; Q,\qquad Q={\alpha+(\beta-\alpha)y-\beta xy\over
  \beta+(\alpha-\beta)x-\alpha xy}\label{hii1}
\end{equation} 
The Miura transformation is
\begin{equation}
x=w\tilde w,\quad u=(\bar w\skew{3.5}\tilde{\bar w})^{-1},\quad
v=w\bar w,\quad y=(\tilde w\skew{3.5}\tilde{\bar
  w})^{-1}\label{hii2}
\end{equation} and substituting into Eq.~(\ref{hii1}) we obtain
\begin{equation}
(\alpha-\beta)(w\tilde w\bar w\skew{3.5}\tilde{\bar
  w}+1)+\beta(w\tilde w+\bar w\skew{3.5}\tilde{\bar w})-\alpha(w\bar
w+\tilde w\skew{3.5}\tilde{\bar w})=0\label{hii3}
\end{equation} 

This equation could be interpreted as a special form of the A2
equation of \cite{AdBoSu07}.  Since Eq.~(\ref{fi3}) has the generic
form of A2 it is convenient to start from this form and put
$q=1+\epsilon\alpha$, $p=1+\epsilon\beta$. Taking the limit
$\epsilon\to0$ we find precisely Eq.~(\ref{hii3}) which establishes
its relation to A2.

\subsection{$H_{\rm III}^{\rm B}$}
We start from
\begin{equation}
u={y\over Q},\qquad v=x\; Q,\qquad Q={\alpha xy+1\over \beta
  xy+1}\label{hiv1}
\end{equation} 
The Miura transformation is now
\begin{equation}
x=p\; w\tilde w,\quad u=-p\; (\bar w\skew{3.5}\tilde{\bar w})^{-1},\quad
v=q\; w\bar w,\quad y=-q\; (\tilde w\skew{3.5}\tilde{\bar
  w})^{-1}\label{hiv2}
\end{equation} 
Eliminating $x,y,v,u$ and introducing $\alpha p^2=1$, $\beta q^2=1$ we
obtain the equation
\begin{equation}
q\; (w\tilde w+\bar w\skew{3.5}\tilde{\bar w})-p\; (w\bar w+\tilde
w\skew{3.5}\tilde{\bar w})=0\label{hiii3}
\end{equation} which is equation H3 of
\cite{AdBoSu07} for the special value $\delta=0$.

We have thus established correspondences via appropriate Miura
transformations, between the two classifications of
\cite{AdBoSu04,PaSuToVe10} and \cite{AdBoSu07}.

Remark: All these constructions can be made
non-autonomous~\cite{SaRaHy07,HiVi11b,GrRa10}.

\section{ A second dynamical interpretation}
\label{yb2rec}

One may view the original lattice in a different way: the bond
variables may be  assigned to the vertices of a finer square lattice, as
shown in Figure~3.

\begin{center}
\setlength{\unitlength}{2947sp}%
\begingroup\makeatletter\ifx\SetFigFont\undefined%
\gdef\SetFigFont#1#2#3#4#5{%
  \reset@font\fontsize{#1}{#2pt}%
  \fontfamily{#3}\fontseries{#4}\fontshape{#5}%
  \selectfont}%
\fi\endgroup%
\begin{picture}(7834,3044)(2989,-6083)
{\color[rgb]{0,0,0}\thinlines
\put(3901,-3361){\circle*{120}}
}%
{\color[rgb]{0,0,0}\put(3901,-4561){\circle*{120}}
}%
{\color[rgb]{0,0,0}\put(4511,-3961){\circle*{120}}
}%
{\color[rgb]{0,0,0}\put(3301,-3961){\circle*{120}}
}%
{\color[rgb]{0,0,0}\put(3301,-5161){\circle*{120}}
}%
{\color[rgb]{0,0,0}\put(4511,-5161){\circle*{120}}
}%
{\color[rgb]{0,0,0}\put(3901,-5771){\circle*{120}}
}%
{\color[rgb]{0,0,0}\put(5101,-5771){\circle*{120}}
}%
{\color[rgb]{0,0,0}\put(5701,-5161){\circle*{120}}
}%
{\color[rgb]{0,0,0}\put(5101,-4561){\circle*{120}}
}%
{\color[rgb]{0,0,0}\put(5701,-3961){\circle*{120}}
}%
{\color[rgb]{0,0,0}\put(5101,-3361){\circle*{120}}
}%

{\color[rgb]{0,0,0}\put(8701,-3361){\circle*{120}}
}%
{\color[rgb]{0,0,0}\put(8701,-4561){\circle*{120}}
}%
{\color[rgb]{0,0,0}\put(9301,-3961){\circle*{120}}
}%
{\color[rgb]{0,0,0}\put(8101,-3961){\circle*{120}}
}%
{\color[rgb]{0,0,0}\put(8101,-5161){\circle*{120}}
}%
{\color[rgb]{0,0,0}\put(9301,-5161){\circle*{120}}
}%
{\color[rgb]{0,0,0}\put(8701,-5761){\circle*{120}}
}%
{\color[rgb]{0,0,0}\put(9901,-5761){\circle*{120}}
}%
{\color[rgb]{0,0,0}\put(10501,-5161){\circle*{120}}
}%
{\color[rgb]{0,0,0}\put(9901,-4561){\circle*{120}}
}%
{\color[rgb]{0,0,0}\put(10501,-3961){\circle*{120}}
}%
{\color[rgb]{0,0,0}\put(9901,-3361){\circle*{120}}
}%
{\color[rgb]{0,0,0}\put(4501,-3061){\line( 0,-1){3000}}
}%
{\color[rgb]{0,0,0}\put(5701,-3061){\line( 0,-1){3000}}
}%
{\color[rgb]{0,0,0}\put(3001,-3361){\line( 1, 0){3000}}
}%
{\color[rgb]{0,0,0}\put(3001,-4561){\line( 1, 0){3000}}
}%
{\color[rgb]{0,0,0}\put(3001,-5761){\line( 1, 0){3000}}
}%
{\color[rgb]{0,0,0}\put(3301,-3061){\line( 0,-1){3000}}
}%
\put(6600,-4561){$\longrightarrow$}
{\color[rgb]{0,0,0}\put(7801,-4261){\line( 1, 1){1200}}
}%
{\color[rgb]{0,0,0}\put(7801,-5461){\line( 1, 1){2400}}
}%
{\color[rgb]{0,0,0}\put(8401,-6061){\line( 1, 1){2400}}
}%
{\color[rgb]{0,0,0}\put(9601,-6061){\line( 1, 1){1200}}
}%
{\color[rgb]{0,0,0}\put(10201,-6061){\line(-1, 1){2400}}
}%
{\color[rgb]{0,0,0}\put(9001,-6061){\line(-1, 1){1200}}
}%
{\color[rgb]{0,0,0}\put(10801,-5461){\line(-1, 1){2400}}
}%
{\color[rgb]{0,0,0}\put(10801,-4261){\line(-1, 1){1200}}
}%
\end{picture}%
\\
Figure~3: A change of point of view
\end{center}

We get a new square lattice whose elementary square cell is, after
rotation of $-{{\pi} \over{4}}$:

\begin{center}
\setlength{\unitlength}{0.0006in}
\begin{picture}(3482,1900)(0,-10)
\put(450,1883){\circle*{120}}
\put(2250,1883){\circle*{120}}
\put(2250,83){\circle*{120}}
\put(450,83){\circle*{120}}
\put(120,1950){\makebox(0,0)[lb]{$v$}}
\put(2380,-10){\makebox(0,0)[lb]{$y$}}
\put(2380,1950){\makebox(0,0)[lb]{$u$}}
\put(110,-5){\makebox(0,0)[lb]{$x$}}
\drawline(450,1883)(450,83)
\drawline(450,1883)(2250,1883)
\drawline(2250,1883)(2250,83)
\drawline(450,83)(2250,83)
\end{picture}
\\
\bigskip
\hskip -1.5 truecm Figure 4: {\em{New  cell}} 
\end{center}

For all the models described in~\cite{AdBoSu04} the conditions
relating $u,v,x,y$ allow to define various rational $2\times 2 $ maps:
they give $u$ and $v$ in terms of $x$ and $y$, but also $x$ and $y$ in
terms of $u$ and $v$, as well as $u$ and $y$ in terms of $v$ and $x$,
or $v$ and $x$ in terms of $u$ and $y$. This is the
``quadrirationality'' property. We obtain in such a way four $2\times
2 $ maps, defined for each elementary cell. These cells form a new
two-dimensional lattice. Notice that there are two kinds of cells in
this lattice. Half of them contained a vertex of the original lattice,
and the other half did not. In the original lattice, we had relations
only for the latter kind of cells.  {\em We will impose the existence
  of relations as above for all the cells, allowing non constant
  parameters}.

This will lead to a possibly non autonomous recurrence of order two.
The variables, denoted by $x_{m,n}$, are assigned to the vertices of
the new lattice. Since on each cell we have four $(m,n)$ dependent
$2\times 2 $ maps, we may, once given initial conditions $x_{0,0}$ and
$x_{0,1}$, calculate $x$ over the whole lattice, provided some
compatibility (zero monodromy around each vertex) condition is
fulfilled.

\begin{center}
\setlength{\unitlength}{3947sp}%
\begingroup\makeatletter\ifx\SetFigFont\undefined%
\gdef\SetFigFont#1#2#3#4#5{%
  \reset@font\fontsize{#1}{#2pt}%
  \fontfamily{#3}\fontseries{#4}\fontshape{#5}%
  \selectfont}%
\fi\endgroup%
\begin{picture}(2865,3129)(2086,-4261)
{\color[rgb]{0,0,0}\thinlines
\put(2401,-1561){\circle*{108}}
}%
{\color[rgb]{0,0,0}\put(3601,-1561){\circle*{108}}
}%
{\color[rgb]{0,0,0}\put(4801,-1561){\circle*{108}}
}%
{\color[rgb]{0,0,0}\put(4801,-2761){\circle*{108}}
}%
{\color[rgb]{1,0,0}\put(3601,-2761){\circle*{108}}
}%
{\color[rgb]{1,0,0}\put(2401,-2761){\circle*{108}}
}%
{\color[rgb]{0,0,0}\put(2401,-3961){\circle*{108}}
}%
{\color[rgb]{0,0,0}\put(3601,-3961){\circle*{108}}
}%
{\color[rgb]{0,0,0}\put(4801,-3961){\circle*{108}}
}%
{\color[rgb]{0,0,0}\put(2401,-1561){\line( 1, 0){1200}}
\put(3601,-1561){\line( 0,-1){1200}}
\put(3601,-2761){\line(-1, 0){1200}}
\put(2401,-2761){\line( 0, 1){1200}}
\put(2401,-1561){\line( 0, 1){  0}}
}%
{\color[rgb]{0,0,0}\put(3601,-1561){\line( 1, 0){1200}}
\put(4801,-1561){\line( 0,-1){1200}}
\put(4801,-2761){\line(-1, 0){1200}}
\put(3601,-2761){\line( 0, 1){1200}}
\put(3601,-1561){\line( 0, 1){  0}}
}%
{\color[rgb]{0,0,0}\put(3601,-2761){\line( 1, 0){1200}}
\put(4801,-2761){\line( 0,-1){1200}}
\put(4801,-3961){\line(-1, 0){1200}}
\put(3601,-3961){\line( 0, 1){1200}}
\put(3601,-2761){\line( 0, 1){  0}}
}%
{\color[rgb]{0,0,0}\put(2401,-2761){\line( 1, 0){1200}}
\put(3601,-2761){\line( 0,-1){1200}}
\put(3601,-3961){\line(-1, 0){1200}}
\put(2401,-3961){\line( 0, 1){1200}}
\put(2401,-2761){\line( 0, 1){  0}}
}%
\thicklines
{\color[rgb]{0,0,0}\put(3001,-2461){\vector( 0, 1){600}}
}%
{\color[rgb]{0,0,0}\put(3901,-2161){\vector( 1, 0){600}}
}%
{\color[rgb]{0,0,0}\put(4201,-3061){\vector( 0,-1){600}}
}%
{\color[rgb]{0,0,0}\put(3301,-3361){\vector(-1, 0){600}}
}%
\put(1700,-2761){\makebox(0,0)[lb]{\smash{{\SetFigFont{12}{14.4}{\rmdefault}{\mddefault}{\updefault}{\color[rgb]{0,0,0}$x_{m,n-1}$}%
}}}}
\put(3666,-2700){\makebox(0,0)[lb]{\smash{{\SetFigFont{12}{14.4}{\rmdefault}{\mddefault}{\updefault}{\color[rgb]{0,0,0}$x_{m,n}$}%
}}}}
\put(2191,-1400){\makebox(0,0)[lb]{\smash{{\SetFigFont{12}{14.4}{\rmdefault}{\mddefault}{\updefault}{\color[rgb]{0,0,0}$x_{m,n-1}$}%
}}}}
\put(3526,-1400){\makebox(0,0)[lb]{\smash{{\SetFigFont{12}{14.4}{\rmdefault}{\mddefault}{\updefault}{\color[rgb]{0,0,0}$x_{m+1,n}$}%
}}}}
\put(4711,-1400){\makebox(0,0)[lb]{\smash{{\SetFigFont{12}{14.4}{\rmdefault}{\mddefault}{\updefault}{\color[rgb]{0,0,0}$x_{m+1,n+1}$}%
}}}}
\put(4936,-2761){\makebox(0,0)[lb]{\smash{{\SetFigFont{12}{14.4}{\rmdefault}{\mddefault}{\updefault}{\color[rgb]{0,0,0}$x_{m,n+1}$}%
}}}}
\put(4786,-4201){\makebox(0,0)[lb]{\smash{{\SetFigFont{12}{14.4}{\rmdefault}{\mddefault}{\updefault}{\color[rgb]{0,0,0}$x_{m-1,n+1}$}%
}}}}
\put(3541,-4201){\makebox(0,0)[lb]{\smash{{\SetFigFont{12}{14.4}{\rmdefault}{\mddefault}{\updefault}{\color[rgb]{0,0,0}$x_{m-1,n}$}%
}}}}
\put(2206,-4186){\makebox(0,0)[lb]{\smash{{\SetFigFont{12}{14.4}{\rmdefault}{\mddefault}{\updefault}{\color[rgb]{0,0,0}$x_{m-1,n-1}$}%
}}}}
\end{picture}%
\\
\bigskip
\hskip .5cm
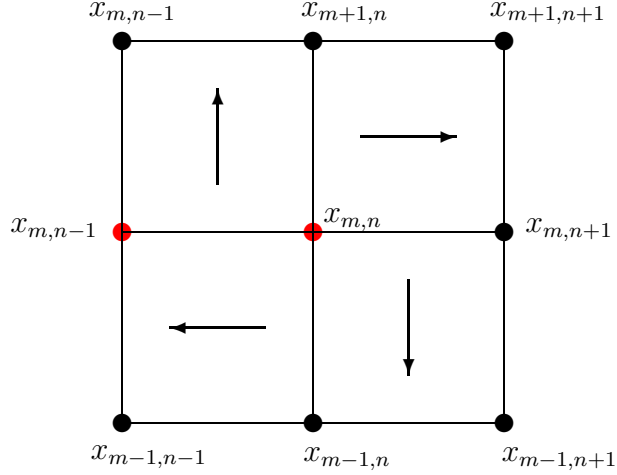
Figure 5: Local compatibility condition around vertex $(m,n)$
\end{center}

We use the coordinates of the lower left corner to label the cells.
Starting from $x_{m,n}$ and $x_{m,n-1}$, we compute $x_{m,n-1}$ and
$x_{m+1,n}$ using the same $F$ and $H$ equations as in the previous
section, with $(m,n)$ dependent set of parameters
$\Gamma_{m,n}$. Going around the vertex $(m,n)$ we calculate the
surrounding values of $x$. This yields after four steps a value for
$x_{m,n-1}$ which has to coincide with the initial condition.

Remark: Doing so we reduce the dimensionality of the system, going
from partial difference equation to ordinary difference equation. The
reason is that we impose relations between variables which were not
related in the original model, actually exactly on half of the cells
of the new lattice.

In what follows, we shall show that it is indeed possible to satisfy
the compatibility constraint.  Moreover the resulting equations are
explicitly solvable ones and given that the compatibility constraints
are also explicitly solvable this allows us to give the solution for
all $x$'s in closed form.

\subsection{ $F_{\rm I}$}
In the case of $F_{\rm I}$,  the starting equations are
\begin{equation*} u=\alpha
y\; P,\qquad v=\beta \; P,\qquad
P={(1-\beta)x+\beta-\alpha+(\alpha-1) y \over
  \beta(1-\alpha)x+(\alpha-\beta)xy+\alpha(\beta-1)y}
\label{gi1}
\end{equation*}
which we rewrite as
\begin{equation*}
x_{m+1,n+1}=\alpha_{m,n}x_{m,n+1}{(1-\beta_{m,n})x_{m,n}+\beta_{m,n}
  -\alpha_{m,n}+(\alpha_{m,n}-1)x_{m,n+1}\over
  \beta_{m,n}(1-\alpha_{m,n})x_{m,n}+(\alpha_{m,n}-\beta_{m,n})x_{m,n}x_{m,n+1}
  + \alpha_{m,n}(\beta_{m,n}-1)x_{m,n+1}}
\end{equation*}
\begin{equation*}
x_{m+1,n}=\beta_{m,n}x_{m,n}
{(1-\beta_{m,n})x_{m,n}+\beta_{m,n}-\alpha_{m,n} + (\alpha_{m,n}-1)
  x_{m,n+1}\over \beta_{m,n}(1-\alpha_{m,n})x_{m,n}+(\alpha_{m,n} -
  \beta_{m,n})x_{m,n}x_{m,n+1}+\alpha_{m,n}(\beta_{m,n}-1)x_{m,n+1}}
\end{equation*}

The compatibility conditions for the $\alpha$ and $\beta$ are
\begin{equation}
\alpha_{m+1,n+1}= \alpha_{m,n},\qquad \beta_{m+1,n} = \beta_{m,n+1}\label{gi4}
\end{equation}
and
\begin{equation}
\alpha_{m+1,n}=\beta_{m,n}\beta_{m,n+1}R,\qquad\beta_{m+1,n+1}=
\alpha_{m,n}\alpha_{m,n+1}R
\label{gi5}
\end{equation}
where
\begin{equation}
R={(\alpha_{m,n}-1)(\alpha_{m,n+1}-1)-(\beta_{m,n}-1)(\beta_{m,n+1}-1)\over
  \alpha_{m,n}\alpha_{m,n+1}(\beta_{m,n}+\beta_{m,n+1}-1) -
  \beta_{m,n}\beta_{m,n+1}(\alpha_{m,n}+\alpha_{m,n+1}-1)}
\label{gi6}
\end{equation}

The set of compatibility equations is itself overdetermined but it
turns out that it is consistent and the solution is periodic in both
$m$ and $n$. We have
$$\alpha_{3m,3n}=\alpha_{3m+1,3n+1}=\alpha_{3m+2,3n+2}=\alpha_0$$
$$\alpha_{3m,3n+1}=\alpha_{3m+1,3n+2}=\alpha_{3m+2,3n}=\alpha_1$$
$$\alpha_{3m,3n+2}=\alpha_{3m+1,3n}=\alpha_{3m+2,3n+1}=\alpha_2$$
$$\beta_{3m,3n}=\beta_{3m+2,3n+1}=\beta_{3m+1,3n+2}=\beta_0\label{gi7}$$
$$\beta_{3m,3n+1}=\beta_{3m+2,3n+2}=\beta_{3m+1,3n}=\beta_1$$
$$\beta_{3m,3n+2}=\beta_{3m+2,3n}=\beta_{3m+1,3n+1}=\beta_2$$ where
$\alpha_2$, $\beta_2$ are given in terms of $\alpha_0$, $\alpha_1$,
$\beta_0$, $\beta_1$ using Eq.~(\ref{gi5}). The solution for
$x_{m,n}$ is equally periodic, with period 3 in both $m$ and $n$
directions.  In this case however all nine $x_{3m+j,3m+k}$ for
$j,k=0,1,2$ are generically different.

\subsection{$F_{\rm II}$}
We start from the equations
\begin{equation}
u={y\over\alpha}P,\qquad v={x\over\beta}P,\qquad P={\alpha x-\beta
  y+\beta-\alpha\over x-y}\label{gii1}
\end{equation}
which we rewrite, introducing $\gamma=\alpha/\beta$, as
\begin{eqnarray}
x_{m+1,n+1} & = & x_{m,n+1}\left({x_{m,n}-x_{m,n+1}/\gamma_{m,n} + 1 -
  1/\gamma_{m,n}\over x_{m,n}-x_{m,n+1}}\right), \nonumber \\
x_{m+1,n} & =& x_{m,n}\left({\gamma_{m,n}x_{m,n}-x_{m,n+1}+\gamma_{m,n}-1\over
  x_{m,n}-x_{m,n+1}}\right)
\label{gii2}
\end{eqnarray}
The compatibility condition is now
\begin{equation*}
\gamma_{m+1,n+1}\gamma_{m+1,n}\gamma_{m,n+1}\gamma_{m,n} -
\gamma_{m+1,n}\gamma_{m,n+1}\gamma_{m,n} -
\gamma_{m+1,n+1}\gamma_{m+1,n}\gamma_{m,n+1} +
\gamma_{m+1,n}+\gamma_{m,n+1}-1=0 
\end{equation*}
which can be rewritten as
\begin{equation}
(1-\gamma_{m+1,n+1})(1-\gamma_{m,n}) =
  (1-1/\gamma_{m+1,n})(1-1/\gamma_{m,n+1})\label{giii4}
\end{equation}
which can be recognized as the linearisable equation of
Hydon-Viallet~(Eq (22) in \cite{HyVi09},  originally
given in terms of the variable $z=1-\gamma$).  The solution to
Eq.~(\ref{giii4}) can be obtained most easily if we introduce an
auxiliary variable $\chi$ and express $\gamma$ as
\begin{equation}
\gamma_{m,n}={(1-\chi_{m+1,n+1})(1-\chi_{m,n})\over(1-\chi_{m+1,n})(1-\chi_{m,n+1})}
\label{gii5}
\end{equation}
with $\chi$ satisfying the equation
\begin{equation}
\chi_{m+1,n+1}\chi_{m,n}=\chi_{m+1,n}\chi_{m,n+1}\label{gii3}
\end{equation}
i.e. $\chi$ is the product of two free functions
$\chi_{m,n}=f(m)g(n)$.  Using this solution we can obtain the general
solution for $x$ as
\begin{equation}
x_{m,n}={af(m)g(n)+bf(m)+cg(n)+a+1\over 1-f(m)g(n)}\label{gii4}
\end{equation}
where $a,b,c$ are constants related by $a(a+1)=bc$.

\subsection{$F_{\rm III}$}
In the case of $F_{\rm III}$, where
\begin{equation}
u={y\over\alpha}P,\qquad v={x\over\beta}P,\qquad P={\alpha x-\beta
  y\over x-y}\label{giii1}
\end{equation}
we introduce the parameter $\gamma=\alpha/\beta$, and  rewrite the
equations as
\begin{equation}
x_{m+1,n+1}=x_{m,n+1}\left({x_{m,n}-x_{m,n+1}/\gamma_{m,n}\over
  x_{m,n}-x_{m,n+1}}\right),\qquad
x_{m+1,n}=x_{m,n}\left({\gamma_{m,n}x_{m,n}-x_{m,n+1}\over
  x_{m,n}-x_{m,n+1}}\right)\label{giii2}
\end{equation}
The compatibility condition is exactly the same as before, namely
$\gamma$ satisfies equation~(\ref{giii4}). In this case, it is more
convenient to express $\gamma$ in terms of an auxiliary variable
$\omega$ as
\begin{equation}
\gamma_{m,n}={\omega_{m+1,n+1}\omega_{m,n}\over\omega_{m+1,n}\omega_{m,n+1}}
\label{giii5}
\end{equation}
and take $\omega$ satisfying the equation
\begin{equation}
\omega_{m+1,n+1}+\omega_{m,n}=\omega_{m+1,n}+\omega_{m,n+1}\label{giii6}
\end{equation}
i.e. $\omega$ is the sum of two free functions
$\omega_{m,n}=f(m)+g(n)$ which leads to the solution for $\gamma$
\begin{equation}
\gamma_{m,n}={(f(m+1)+g(n+1))(f(m)+g(n))
  \over(f(m+1)+g(n))(f(m)+g(n+1))} \label{giii7}
\end{equation}
The general solution for $x$ can now be given
\begin{equation}
x_{m,n}=a{(f(m)+b)(g(n)-b)\over f(m)+g(n)}\label{giii8}
\end{equation}
where $a,b$ are constants.

\subsection{ $F_{\rm IV}$}
The defining equations
\begin{equation}
u=y\; P,\qquad v=x\; P,\qquad P=1-{\alpha-\beta\over x-y}\label{giv1}
\end{equation}
are rewritten as:
\begin{equation}
x_{m+1,n+1}=x_{m,n+1}\left(1-{\gamma_{m,n}\over
  x_{m,n}-x_{m,n+1}}\right),\qquad
x_{m+1,n}=x_{m,n}\left(1-{\gamma_{m,n}\over
  x_{m,n}-x_{m,n+1}}\right)\label{giv2}
\end{equation}
where $\gamma=\alpha-\beta$.

The compatibility condition in this case turns out to be
\begin{equation}
\gamma_{m+1,n+1}\gamma_{m,n}=\gamma_{m+1,n}\gamma_{m,n+1}\label{gv3}
\end{equation}
the solution of which is $\gamma_{m,n}=\phi(m)\psi(n)$, where $\phi$
and $\psi$ are free functions of their argument. At this point is is
convenient to introduce two functions $f(m)$ and $g(n)$ such that
$f(m+1)-f(m)=\phi(m)$ and $g(n+1)-g(n)=\psi(n)$.
 The general
solution here can be written as
\begin{equation}
x_{m,n}=(f(m)+a)(g(n)+b)\label{giv3}
\end{equation}
where $a,b$ are constants.

\subsection{ $F_{\rm V}$}
The defining equations are
\begin{equation}
u=y+P,\qquad v=x+P,\qquad P={\gamma\over x-y}\label{gv1}
\end{equation}
with the same definition of $\gamma$ as before, and  we rewrite them as
\begin{equation}
x_{m+1,n+1}=x_{m,n+1}+{\gamma_{m,n}\over x_{m,n}-x_{m,n+1}},\qquad
x_{m+1,n}=x_{m,n}+{\gamma_{m,n}\over x_{m,n}-x_{m,n+1}}\label{gv2}
\end{equation}

The compatibility condition in this case turns out to be the same as
in the previous case, namely Eq.~(\ref{gv3}). Using the same functions $f$
and $g$ as above, it is straightforward to verify that
\begin{equation}
x_{m,n}=af(m)+{g(n)\over a}+b\label{gv4}
\end{equation}
is the general solution of system (\ref{gv2}), where $a,b$ are
constants.

Next we turn to the study of the $H$ family.

\subsection{ $H_{\rm I}$}
The first case to examine is $H_{\rm I}$
\begin{equation}
u = yQ^{-1},\qquad v=xQ,\qquad Q = {(1-\beta)xy + (\beta-\alpha)y +
  \beta(\alpha-1)\over(1-\alpha)xy+(\alpha-\beta)x+\alpha(\beta-1)}
\label{ki1}
\end{equation}
which we rewrite, after a translation $\alpha=\gamma+1$, $\beta=\delta+
1$, as
\begin{equation}x_{m+1,n+1}=x_{m,n+1}{\gamma_{m,n}x_{m,n}x_{m,n+1}-(\gamma_{m,n} - 
\delta_{m,n})x_{m,n}-\delta_{m,n}(\gamma_{m,n}+1)\over
\delta_{m,n}x_{m,n}x
_{m,n+1}+(\gamma_{m,n}-\delta_{m,n})x_{m,n+1}-\gamma_{m,n}(\delta_{m,n}+1)}
\label{ki2}
\end{equation}
\begin{equation}x_{m+1,n}=x_{m,n}
{\delta_{m,n}x_{m,n}x_{m,n+1}+(\gamma_{m,n}-\delta_{m,n})x_{m,n+1} -
  \gamma_{m,n}(\delta_{m,n}+1)\over
  \gamma_{m,n}x_{m,n}x_{m,n+1}-(\gamma_{m,n} -
  \delta_{m,n})x_{m,n}-\delta_{m,n}(\gamma_{m,n}+1)}
\label{ki3}
\end{equation}
In this case the two parameters $\gamma$ and $\delta$ cannot be
combined into a single parameter. The compatibility condition turns
out to be
\begin{equation}\gamma_{m+1,n}=\delta_{m,n}, \qquad \delta_{m+1,n}=\gamma_{m,n}
\label{ki4}
\end{equation}
which means that $\gamma$ and $\delta$ can be expressed in terms of
two free functions $f,g$ of $n$ alone:
$\gamma_{2m,n}=\delta_{2m+1,n}=f(n)$,
$\gamma_{2m+1,n}=\delta_{2m,n}=g(n)$. The solution of
Eqs.~(\ref{ki2}), (\ref{ki3}) is also periodic in the $m$-direction:
$x_{m+2,n}=x_{m,n}$. Thus the equation becomes a one-dimensional one
for the variable $x$ at given parity of $m$. Once we have obtained the
solution for one parity the solution at the other parity is obtained
through Eq.~(\ref{ki2}) or (\ref{ki3}). Taking the index $m$ to be
even we obtain the equation for $x$
\begin{eqnarray} 
&& (g(n) g(n-1) - f(n) f(n-1)) x_{n-1} x_n x_{n+1} + \nonumber \\ & +
  & g(n)(f(n-1)-g(n-1))\; x_nx_{n+1}+f(n-1)(f(n)-g(n))\; x_nx_{n-1} +
  \nonumber \\ & + & f(n-1)(g(n-1)+1)(f(n)-g(n))\; x_{n+1} +
  g(n)(f(n)+1)(f(n-1)-g(n-1)) \; x_{n-1} \nonumber + \\ & + &
  g(n)g(n-1)(f(n)+f(n-1)+1)-f(n)f(n-1)(g(n)+g(n-1)+1) = 0
\label{ki5}
\end{eqnarray}
where we have dropped the even index $m$. (In fact the equation for
odd-$m$ is the same up to the permutation of $f$ and $g$). Equation
(\ref{ki5}) is linearisable. A study of the growth of the degrees of
the iterates on an initial condition, using algebraic entropy
techniques, leads to the following sequence of degrees:
0,1,1,2,2,3,3,4,4,5,~$\dots$. This suggest that the mapping defined by
(\ref{ki5}) is of Gambier type~\cite{GrRa95} and this turns out to be indeed the
case. We find the invariant
\begin{equation}
K=(-1)^n {f(n)(x_n+1)(x_{n+1}-1) + g(n)(x_n-1)(x_{n+1}+1) - 2f(n)g(n)
  \over (f(n)-g(n))(x_n-1)(x_{n+1}-1)}
\label{ki6}
\end{equation}
which allows to reduce Eq.~(\ref{ki5}) to a homographic mapping.

\subsection{ $H_{\rm II}$}
The defining equations 
\begin{equation}
u=y\; Q^{-1},\qquad v=x\; Q,\qquad Q={\alpha+(\beta-\alpha)y-\beta
  xy\over \beta+(\alpha-\beta)x-\alpha xy}\label{kii1}
\end{equation}
 can be cast into the form
\begin{equation}
x_{m+1,n+1}=x_{m,n+1}{1+(\gamma_{m,n}-1)x_{m,n}-\gamma_{m,n}x_{m,n}x_{m,n+1}\over
  \gamma_{m,n}+(1-\gamma_{m,n})x_{m,n+1}-x_{m,n}x_{m,n+1}}\label{kii2}
\end{equation}
\begin{equation}
x_{m+1,n}=x_{m,n}{\gamma_{m,n}+(1-\gamma_{m,n})x_{m,n+1}-x_{m,n}x_{m,n+1}\over
  1+(\gamma_{m,n}-1)x_{m,n}-\gamma_{m,n}x_{m,n}x_{m,n+1}}\label{kii3}
\end{equation}
where $\gamma=\alpha/\beta$.  It turns out that the compatibility
condition for $\gamma$ is precisely Eq.~(\ref{giii4}) i.e. the
Hydon-Viallet equation. We give the solution for $\gamma$ in the same
form as in the case of $F_{\rm II}$, namely
\begin{equation}
\gamma_{m,n}={(1-f(n+1)g(m+1))(1-f(n)g(m))\over(1-f(n+1)g(m))(1-f(n)g(m+1))}
\label{kii4}
\end{equation}
We can now give the solution for $x$ which turns out to possess an
even-odd freedom
\begin{equation}
x_{m,n}={(1+a_nf(n))(g(m)+a_{n+1})\over(1+a_{n+1}f(n))(g(m)+a_n)}\label{kii5}
\end{equation}
where $a_n$ is a period-two constant, i.e. $a_{n+2}=a_n$.

\subsection{$H_{\rm III}^{\rm B}$}
The set of equations  $H_{\rm III}^{\rm B}$
\begin{equation}
u={y\over Q},\qquad v=x\; Q,\qquad Q={\alpha xy+1\over \beta
  xy+1}\label{kiv1}
\end{equation}
yield
\begin{equation}
x_{m+1,n+1}=x_{m,n+1}{\beta_{m,n}x_{m,n}x_{m,n+1}+1\over
  \alpha_{m,n}x_{m,n}x_{m,n+1}+1},\qquad
x_{m+1,n}=x_{m,n}{\alpha_{m,n}x_{m,n}x_{m,n+1}+1\over
  \beta_{m,n}x_{m,n}x_{m,n+1}+1}\label{kiv2}
\end{equation}
The latter can be written in more compact form if we introduce
$\alpha=\gamma+\delta$ and $\beta=\gamma-\delta$. We obtain
\begin{equation}
\delta_{m,n}\delta_{m+1,n+1}=\delta_{m,n+1}\delta_{m+1,n}\label{kiv3}
\end{equation}
\begin{equation}
(\gamma_{m+1,n}-\gamma_{m,n})\delta_{m+1,n+1} +
  (\gamma_{m+1,n+1}-\gamma_{m,n+1})\delta_{m+1,n}=0\label{kiv4}
\end{equation}
which can be readily integrated to
\begin{equation}
\alpha_{m,n}=f(n)(h(n)+(-1)^nk(m)+g(m))\label{kiv5}
\end{equation}
\begin{equation}
\beta_{m,n}=f(n)(h(n)+(-1)^nk(m)-g(m))\label{kiv6}
\end{equation}
Next we remark that from Eq.~(\ref{kiv2}) we have
\begin{equation}
x_{m,n}x_{m,n+1}=x_{m+1,n}x_{m+1,n+1}\equiv \phi(n)\label{kiv7}
\end{equation}
where $\phi$ is, obviously, independent of $m$. Given the expressions
of $\alpha$ and $\beta$ we can obtain the value of
$1/\phi=f(n)(-h(n)+(-1)^nc)$, where $c$ is an integration constant,
the value of which is fixed by the initial conditions. The values of
$x$ can now be obtained from the recursion
\begin{equation}
x_{m,n+2}={\phi(n+1)\over\phi(n)}x_{m,n}\label{kiv8}
\end{equation}
Using the expression for $\phi$ we can, starting from some initial
condition $x_{0,0},x_{0,1}$ obtain all $x_{0,n}$ for all $n$.  Once
the $x_{0,n}$ are obtained we can compute $x_{m,n}$ by
\begin{equation}
x_{m+1,2n}={c-k(m)-g(m)\over c-k(m)+g(m)}x_{m,2n}\label{kiv9}
\end{equation}
\begin{equation}
x_{m+1,2n+1}={c-k(m)+g(m)\over c-k(m)-g(m)}x_{m,2n+1}\label{kiv10}
\end{equation}
Thus the solution to Eq.~(\ref{kiv2}) can be expressed as a double
semi-infinite product.

\section{Conclusion}
Integrable discrete equations appear in an innumerable number of
avatars, and we have exhibited some relations between such forms.

The relations we have described enter the general problem of
classifying integrable discrete equations. One should nevertheless
emphasize that the correspondences we have established are not
invertible.  They consequently do not define equivalences between
models.

This is reminiscent of what happens when one goes from a discrete
equation to its ``potential form'', or when one performs reductions by
symmetries.  These are other types of transformations establishing
relations between different integrable systems, and the links between
these various correspondences and the one we presented here still need
to be explored.

{\bf Acknowledgment:} We would like to thank Jarmo Hietarinta for
stimulating discussions in the early stage of this work.


\end{document}